\documentclass[11pt]{article}

\usepackage[margin=1in]{geometry}
\usepackage{amsmath,amssymb,amsthm,mathtools}
\usepackage{bm}
\usepackage{graphicx}
\usepackage{booktabs}
\usepackage{siunitx}
\usepackage{enumitem}
\usepackage[numbers]{natbib}
\usepackage{subcaption}
\usepackage{microtype}
\usepackage{algorithm}
\usepackage{algpseudocode}
\usepackage{float} 
\usepackage[hidelinks]{hyperref} 

\newtheorem{theorem}{Theorem}
\newtheorem{proposition}{Proposition}
\newtheorem{lemma}{Lemma}

\newtheorem{remark}{Remark}

\newcommand{\KL}[2]{D_{\mathrm{KL}}\!\left(#1\middle\|#2\right)}
\newcommand{\E}{\mathbb{E}}
\newcommand{\Var}{\mathrm{Var}}
\newcommand{\Cov}{\mathrm{Cov}}
\newcommand{\R}{\mathbb{R}}
\newcommand{\1}{\mathbf{1}}
\newcommand{\bb}{\mathbf{b}}
\newcommand{\ww}{\mathbf{w}}
\newcommand{\xx}{\mathbf{x}}
\newcommand{\XX}{\mathbf{X}}

\title{\textbf{Entropy-Guided Multiplicative Updates:}\\
KL Projections for Multi-Factor Target Exposures}
\author{Yimeng Qiu}
\date{2025/06/18}

\begin{document}
\maketitle

\begin{abstract}
We develop \emph{Entropy-Guided Multiplicative Updates} (EGMU), a convex optimization framework for constructing multi-factor target-exposure portfolios by minimizing Kullback--Leibler (KL) divergence from a benchmark subject to linear factor constraints.
Our contributions are theoretical and algorithmic. 
(\emph{i}) We formalize feasibility and uniqueness: with strictly positive benchmark and feasible targets in the convex hull of exposures, the solution is unique and strictly positive. 
(\emph{ii}) We derive the dual concave program with gradient $t-\E_{w(\theta)}[x]$ and Hessian $-\Cov_{w(\theta)}(x)$, and give a precise sensitivity formula $\partial\theta^*/\partial t=\Cov_{w^*}(x)^{-1}$ and $\partial w^*/\partial t=\mathrm{diag}(w^*) (X-\1\mu^\top)\Cov_{w^*}(x)^{-1}$. 
(\emph{iii}) We present two provably convergent solvers: a damped \emph{dual Newton} method with global convergence and local quadratic rate, and a \emph{KL-projection} scheme based on IPF/Bregman--Dykstra for equalities and inequalities. 
(\emph{iv}) We further \textbf{generalize EGMU} with \emph{elastic targets} (strongly concave dual) and \emph{robust target sets} (support-function dual), and introduce a \emph{path-following ODE} for solution trajectories, all reusing the same dual-moment structure and solved via Newton or proximal-gradient schemes.
(\emph{v}) We detail numerically stable and scalable implementations (LogSumExp, covariance regularization, half-space KL-projections). 
We emphasize theory and reproducible algorithms; empirical benchmarking is optional.
\end{abstract}


\section{Introduction}
Rules-based multi-factor portfolios seek specified exposures (Value, Momentum, Quality, Low Volatility, etc.).
Heuristic sequential ``tilts'' lack a single global objective and are order-dependent.
Quadratic exposure-matching solves a different closeness metric and often needs explicit regularization and a risk model.

We pose \emph{Entropy-Guided Multiplicative Updates} (EGMU): minimize $\KL{\ww}{\bb}$ over the simplex under linear exposure constraints. This information projection is classical and yields exponential-family solutions and convex duality structure \citep{csiszar1975i,cover2006elements}. In portfolio engineering it parallels Entropy Pooling \citep{meucci2008fully}. Our focus is to provide a rigorous, self-contained treatment tailored to target-exposure construction: feasibility/uniqueness, sensitivity, and provably convergent algorithms for equality and inequality constraints. We also give implementable pseudo-code with stability safeguards. Our generalized variants---elastic/robust targets and solution paths---remain within the same dual-moment framework.

\paragraph{Notation.}
Let $N$ be the number of assets, $K$ factors. Benchmark $\bb\in\Delta^N:=\{\ww\!\in\!\R_{\ge0}^N:\1^\top\ww=1\}$ has strictly positive entries ($b_i>0$). Exposure matrix $\XX\in\R^{N\times K}$ has rows $\xx_i^\top$. Targets $t\in\R^K$. Expectations $\E_w[\cdot]$ are under the discrete distribution $w$ on $\{1,\dots,N\}$.

\paragraph{Operators and shorthand.}
For a vector $v$, $\mathrm{normalize}(v):=v/(\1^\top v)$ projects $v$ onto the simplex ray.
Elementwise product/division are denoted by $\odot$ and $\oslash$.
We write $\Pi^{\mathrm{KL}}_{\mathcal{C}}(u)$ for the (unique) KL projection of $u$ onto a closed convex set $\mathcal{C}\subseteq\Delta^N$.
We use the LogSumExp trick $\log\!\sum_i b_i e^{s_i}=\log\!\sum_i b_i e^{s_i-m}+m$ with $m=\max_i s_i$.
We adopt $\E_{w}[x]=\sum_i w_i x_i$ and $\Cov_w(x)=\sum_i w_i(x_i-\mu)(x_i-\mu)^\top$ with $\mu=\E_w[x]$.

\paragraph{Contributions at a glance.}
\begin{itemize}[leftmargin=1.2em,topsep=2pt,itemsep=1pt]
\item \textbf{KL-based target-exposure construction}: existence/uniqueness and the exponential-family solution with covariance Hessian.
\item \textbf{Algorithms with guarantees}: a damped dual Newton method and Bregman projection schemes (IPF/Dykstra) for equalities and inequalities.
\item \textbf{Generalizations with shared dual moments}: elastic targets (strongly concave dual) and robust target sets (support-function dual) with a proximal-gradient solver.
\item \textbf{Sensitivity and paths}: closed-form sensitivities and a homotopy ODE to trace optimal solutions along target directions.
\end{itemize}

\section{Problem, Feasibility, and Geometry}

\subsection{KL-Minimization with Linear Constraints}
We study
\begin{equation}
\label{eq:primal}
\min_{\ww\in\Delta^N}\ \KL{\ww}{\bb}\quad
\text{s.t.}\quad \XX^\top\ww = t,\qquad A\ww\le c.
\end{equation}
The objective is strictly convex on the relative interior of $\Delta^N$; the feasible set is convex.

\subsection{Feasibility and Strict Positivity}
\begin{proposition}[Feasibility and strict positivity]
\label{prop:feasible}
If only equality constraints $\XX^\top \ww=t$ are present, feasibility holds iff $t\in\mathrm{conv}\{ \xx_i\}_{i=1}^N$. If $t$ lies in the relative interior of $\mathrm{conv}\{\xx_i\}$, the unique optimizer satisfies $w_i^\star>0$ for all $i$. With additional linear inequalities $A\ww\le c$, feasibility remains a convex polytope; infeasibility admits a Farkas-type certificate. 
\end{proposition}

\begin{remark}[Intercept factor, linear dependence, and gauge fixing]
\label{rem:intercept}
If $\XX$ contains a constant (intercept) column $\1$, then the budget constraint $\1^\top \ww=1$ is linearly redundant with that column, making the dual variable non-unique and the covariance $\Cov_{w(\theta)}(x)$ singular along the intercept direction.
In practice, \emph{remove the intercept column and keep the budget}, or equivalently keep the intercept but \emph{fix its dual component to zero} (gauge fixing).
Numerically, this avoids singular normal equations and yields a well-posed Newton step on the $K{-}1$ dimensional exposure subspace.
\end{remark}

\begin{lemma}[Column-shift (translation) invariance]
\label{lem:shift}
For any $d\in\R^K$, define $X':=X-\1 d^\top$ and $t':=t-d$. Then the equality-feasible sets coincide:
\[
\{w\in\Delta^N:\ X^\top w=t\}\;=\;\{w\in\Delta^N:\ X'^\top w=t'\}.
\]
\emph{Proof.} One line: $X'^\top w=(X-\1 d^\top)^\top w=X^\top w-d(\1^\top w)=t-d$ since $\1^\top w=1$.
\end{lemma}

\section{Duality and Exponential-Family Form}

\subsection{Exponential Tilt (Equality Case)}
With only $\XX^\top\ww=t$, the KKT conditions give the exponential-family solution
\begin{equation}
\label{eq:exp-solution}
w_i(\theta)=\frac{b_i\exp(\theta^\top \xx_i)}{\sum_{j} b_j \exp(\theta^\top \xx_j)}.
\end{equation}
The dual concave objective reads
\begin{equation}
\label{eq:dual}
L(\theta)=\theta^\top t - \log\!\Big(\sum_i b_i e^{\theta^\top \xx_i}\Big),
\end{equation}
with
\[
\nabla L(\theta)=t-\E_{w(\theta)}[\xx],\qquad
\nabla^2 L(\theta)=-\Cov_{w(\theta)}(\xx).
\]
Strict concavity holds on the span where $\Cov_{w(\theta)}(\xx)\succ 0$ \citep{wainwright2008graphical}.

\subsection{Sensitivity to Targets}
Let $\theta^\star$ maximize $L$, $\mu=\E_{w^\star}[\xx]$, and $\Sigma=\Cov_{w^\star}(\xx)$. Then
\begin{equation}
\label{eq:sensitivity}
\frac{\partial \theta^\star}{\partial t}=\Sigma^{-1},\qquad
\frac{\partial w_i^\star}{\partial t}=w_i^\star(\xx_i-\mu)^\top \Sigma^{-1}.
\end{equation}

\subsection{Elastic Targets (Soft Penalty): Dual, Uniqueness, and Sensitivity}
\label{sec:elastic}
Consider the elastic objective
\[
\min_{w\in\Delta^N}\ \KL{w}{b}+\frac{\lambda_{\mathrm{soft}}}{2}\,\|X^\top w-t\|_2^2.
\]
Its Fenchel dual is
\[
\max_{\theta\in\R^K}\quad L_{\mathrm{el}}(\theta):=\theta^\top t-\log\!\sum_i b_i e^{\theta^\top x_i}-\frac{1}{2\lambda_{\mathrm{soft}}}\|\theta\|_2^2,
\]
so the optimizer is unique and the primal solution remains the exponential tilt $w_i\propto b_i e^{\theta^{\star\top}x_i}$. The gradient/Hessian of $L_{\mathrm{el}}$ are
\[
\nabla L_{\mathrm{el}}(\theta)=t-\E_{w(\theta)}[x]-\tfrac{1}{\lambda_{\mathrm{soft}}}\theta,\qquad
\nabla^2 L_{\mathrm{el}}(\theta)=-\Cov_{w(\theta)}(x)-\tfrac{1}{\lambda_{\mathrm{soft}}}I.
\]
\begin{theorem}[Elastic sensitivity]
\label{thm:elastic-sens}
At $\theta^\star_{\mathrm{el}}$, we have
\[
\frac{\partial \theta^\star_{\mathrm{el}}}{\partial t}=\big(\Sigma+\tfrac{1}{\lambda_{\mathrm{soft}}}I\big)^{-1},\qquad
\frac{\partial w^\star}{\partial t}=\mathrm{diag}(w^\star)\,(X-\1\mu^\top)\,\big(\Sigma+\tfrac{1}{\lambda_{\mathrm{soft}}}I\big)^{-1}.
\]
\end{theorem}

\subsection{Robust Target Sets via Support Functions}
\label{sec:robust}
Relax the equality to a convex set: $X^\top w\in t_0+\mathcal U$ for a closed, convex, centrally-symmetric set $\mathcal U\subset\R^K$. Then
\[
\min_{w\in\Delta^N}\ \KL{w}{b}+\iota_{\,t_0+\mathcal U}(X^\top w)
\quad\Longleftrightarrow\quad
\max_{\theta\in\R^K}\ L_{\mathrm{rob}}(\theta):=\sigma_{t_0+\mathcal U}(\theta)-\log\!\sum_i b_i e^{\theta^\top x_i},
\]
where $\sigma_S(\theta)$ is the support function. In particular,
\[
\mathcal U=\{u:\|u\|_2\le\rho\}\Rightarrow \sigma_{t_0+\mathcal U}(\theta)=\theta^\top t_0+\rho\|\theta\|_2;\quad
\mathcal U=\{u:\|u\|_\infty\le\rho\}\Rightarrow \sigma_{t_0+\mathcal U}(\theta)=\theta^\top t_0+\rho\|\theta\|_1.
\]
The primal optimizer keeps the exponential tilt $w_i\propto b_i e^{\theta^{\star\top}x_i}$.

\section{Algorithms}

\subsection{EGMU-Newton: Damped Dual Newton Ascent (Equality Core)}
We solve \eqref{eq:dual} via Newton steps with backtracking. Each iteration forms $\mu=\E_{w(\theta)}[\xx]$ and $\Sigma=\Cov_{w(\theta)}(\xx)$ in $O(NK)$ and $O(NK^2)$, and solves $\Sigma\,\Delta=g$ with $g=t-\mu$.

\begin{algorithm}[H]
\caption{EGMU-Newton (Equality Case, LogSumExp-stable)}
\label{alg:newton}
\begin{algorithmic}[1]
\State \textbf{Input}: $b\in\Delta^N$, $X\in\R^{N\times K}$, target $t$, tol $\varepsilon$, ridge $\delta\ge0$
\State Initialize $\theta\leftarrow 0$
\While{$\|\nabla L(\theta)\|_2>\varepsilon$}
  \State \textbf{Scores:} $s_i\leftarrow \theta^\top x_i$;\quad $m\leftarrow \max_i s_i$
  \State \textbf{Log-sum-exp:} $\log Z \leftarrow \log\!\sum_i b_i \exp(s_i-m)+m$
  \State \textbf{Weights:} $w_i\leftarrow b_i \exp(s_i-\log Z)$
  \State \textbf{Moments:} $\mu\leftarrow X^\top w$;\quad $g\leftarrow t-\mu$
  \State \textbf{Covariance:} $\Sigma\leftarrow \sum_i w_i (x_i-\mu)(x_i-\mu)^\top$
  \State \textbf{Solve:} $(\Sigma+\delta I)\Delta = g$ \Comment{Cholesky; $\delta$ only if needed}
  \State \textbf{Line search:} Armijo backtracking with parameters $(c,\beta)$
  \State $\theta\leftarrow \theta+\alpha\Delta$
\EndWhile
\State \textbf{Return} $w(\theta)$ via \eqref{eq:exp-solution}
\end{algorithmic}
\end{algorithm}

\paragraph{Line-search parameters.}
Choose $c\in(10^{-6},10^{-1})$ and $\beta\in(0,1)$ (e.g., $\beta=0.5$); pick the largest $\alpha=\beta^m$ such that
$L(\theta+\alpha\Delta)\ge L(\theta)+c\,\alpha\,g^\top\Delta$.

\paragraph{Elastic variant (R1).}
For $L_{\mathrm{el}}(\theta)$, reuse Algorithm~\ref{alg:newton} with
\[
g\leftarrow t-\mu-\tfrac{1}{\lambda_{\mathrm{soft}}}\theta,\qquad
\Sigma\leftarrow \Sigma+\tfrac{1}{\lambda_{\mathrm{soft}}}I.
\]
This preserves global convergence and improves conditioning via the $I/\lambda_{\mathrm{soft}}$ term.

\subsection{KL-Projections for Equalities: IPF / One-Dimensional Solves}
For a single equality $a^\top w=\tau$, the KL projection of $u$ onto that hyperplane has closed form
\[
w(\alpha)\ \propto\ u\odot \exp(\alpha a),\quad \text{with }\ \phi(\alpha):=a^\top w(\alpha)-\tau=0,
\]
where $\phi$ is strictly monotone since $\phi'(\alpha)=\Var_{w(\alpha)}(a)>0$ unless $a$ is degenerate. Root $\alpha$ is found by bisection/Brent in $O(N)$.
Cycling over $k=1,\dots,K$ yields IPF/GIS; it converges to the KL minimizer under feasibility \citep{csiszar1975i,darroch1972generalized}.

\begin{algorithm}[H]
\caption{EGMU-IPF (Equalities via KL One-Dimensional Projections)}
\label{alg:ipf}
\begin{algorithmic}[1]
\State \textbf{Input}: prior $u\in\Delta^N$, constraints $\{(a_k,\tau_k)\}_{k=1}^K$, tol $\varepsilon$
\State $w\leftarrow u$
\Repeat
  \For{$k=1$ \textbf{to} $K$}
    \State Find $\alpha$ s.t.\ $a_k^\top \big(\mathrm{normalize}(w\odot e^{\alpha a_k})\big)=\tau_k$ \Comment{bisection/Brent}
    \State $w \leftarrow \mathrm{normalize}(w\odot e^{\alpha a_k})$
  \EndFor
\Until{$\max_k |a_k^\top w-\tau_k|\le \varepsilon$}
\State \textbf{Return} $w$
\end{algorithmic}
\end{algorithm}

\subsection{KL-Projections for Inequalities: Bregman--Dykstra}
For a half-space $\mathcal{H}=\{w: a^\top w\le \tau\}$, the KL projection of $u$ onto $\mathcal{H}$ is either $u$ (if feasible) or $w(\lambda)\propto u\odot e^{-\lambda a}$ with $\lambda\ge 0$ chosen so that $a^\top w(\lambda)=\tau$.
Bregman--Dykstra cycles projections onto $\{\mathcal{C}_j\}$ with correction terms $\{q_j\}$ and converges to the KL-projection onto $\cap_j\mathcal{C}_j$ \citep{bauschke2000dykstra}.
Moreover, since $\tfrac{d}{d\lambda}\,a^\top w(\lambda)=-\Var_{w(\lambda)}(a)\le 0$, the residual $a^\top w(\lambda)-\tau$ is strictly decreasing in $\lambda$ (unless $a$ is degenerate), so the one-dimensional root-finding is robust and unimodal.

\begin{algorithm}[H]
\caption{EGMU-Projection (Inequalities via KL Bregman--Dykstra)}
\label{alg:dykstra}
\begin{algorithmic}[1]
\State \textbf{Input}: prior $u\in\Delta^N$, sets $\{\mathcal{C}_j\}_{j=1}^J$ (equalities/half-spaces), tol $\varepsilon$
\State $w\leftarrow u$;\quad $q_j\leftarrow \1$ for all $j$
\Repeat
  \For{$j=1$ \textbf{to} $J$}
    \State $y \leftarrow \mathrm{normalize}(w\odot q_j)$
    \State $z \leftarrow \Pi^{\mathrm{KL}}_{\mathcal{C}_j}(y)$ \Comment{closed-form or 1-D solve as above}
    \State $q_j \leftarrow (w\odot q_j)\oslash z$ \Comment{elementwise}
    \State $w \leftarrow z$
  \EndFor
\Until{constraint violations $\le \varepsilon$}
\State \textbf{Return} $w$
\end{algorithmic}
\end{algorithm}

\subsection{EGMU-ProxGrad (Robust Dual, R2)}
\label{alg:proxgrad}
For $L_{\mathrm{rob}}(\theta)=\underbrace{\theta^\top t_0-\log\!\sum_i b_i e^{\theta^\top x_i}}_{\text{smooth concave }f(\theta)}\ +\ \underbrace{\sigma_{\mathcal U}(\theta)}_{\text{convex}},
$ apply proximal gradient ascent
\[
\theta^{+}\;=\;\mathrm{prox}_{\eta\,\sigma_{\mathcal U}}\big(\theta+\eta\,\nabla f(\theta)\big),
\quad\text{with}\quad \nabla f(\theta)=t_0-\E_{w(\theta)}[x].
\]
By Moreau's identity, $\mathrm{prox}_{\eta\,\sigma_{\mathcal U}}(z)=z-\eta\,\Pi_{\mathcal U}(z/\eta)$ (see, e.g., \citealp{bauschke2011convex}), where $\Pi_{\mathcal U}$ is the Euclidean projection onto $\mathcal U$ (closed forms: $\ell_2$ ball $\Rightarrow$ radial shrink; $\ell_\infty$ box $\Rightarrow$ coordinatewise clip).

\begin{algorithm}[H]
\caption{EGMU-ProxGrad (Robust Dual with $\ell_2/\ell_\infty$ target sets)}
\label{alg:prox}
\begin{algorithmic}[1]
\State \textbf{Input}: $b,X,t_0$, convex $\mathcal U$ (e.g., $\ell_2$ ball radius $\rho$ or $\ell_\infty$ box), step $\eta>0$, tol $\varepsilon$
\State Initialize $\theta\leftarrow 0$
\Repeat
  \State $w_i \propto b_i e^{\theta^\top x_i}$; normalize $w$
  \State $g \leftarrow t_0 - X^\top w$ \Comment{$=\nabla f(\theta)$}
  \State $z \leftarrow \theta + \eta g$
  \State \textbf{Prox:}\quad $\theta \leftarrow z - \eta\,\Pi_{\mathcal U}(z/\eta)$
\Until{$\|\nabla f(\theta)-u\|_2\le\varepsilon$ for some $u\in\partial\sigma_{\mathcal U}(\theta)$}
\State \textbf{Return} $w(\theta)$
\end{algorithmic}
\end{algorithm}

\paragraph{When to use which solver.}
Use \textbf{Algorithm~\ref{alg:newton}} for fast equality matching (small $K$, large $N$). Use the \textbf{elastic variant} in \S\ref{sec:elastic} when exact feasibility is difficult or undesirable. Use \textbf{Algorithm~\ref{alg:prox}} for robust target sets (\(\ell_2/\ell_\infty\)) or when you want feasibility-by-construction via projections.

\subsection{Path-Following via Sensitivity ODE (Module C)}
\label{sec:path}
For a target path $t(\lambda)=t_0+\lambda\Delta$, the optimal dual parameter satisfies the ODE
\[
\frac{d\theta(\lambda)}{d\lambda}=
\Big(\Sigma(\theta(\lambda))+\tfrac{1}{\lambda_{\mathrm{soft}}}I\Big)^{-1}\Delta,\qquad
\theta(0)=\theta^\star(t_0),\ \lambda\in[0,1],
\]
with $\lambda_{\mathrm{soft}}=\infty$ for the equality case. The path is unique under $\Sigma\succeq mI$ and locally Lipschitz Hessian; for robust sets it is piecewise smooth (kinks when the active face of $\mathcal U$ changes).

\begin{algorithm}[H]
\caption{EGMU-Path (Homotopy Integrator)}
\label{alg:path}
\begin{algorithmic}[1]
\State \textbf{Input}: $b,X,t_0,\Delta$, (optional) $\lambda_{\mathrm{soft}}$, step $h>0$
\State Initialize $\theta\leftarrow \theta^\star(t_0)$ (or $0$)
\For{$\lambda=0$ \textbf{to} $1$ \textbf{step} $h$}
  \State $w_i \propto b_i e^{\theta^\top x_i}$; normalize $w$
  \State $\mu\leftarrow X^\top w$;\quad $\Sigma\leftarrow \sum_i w_i(x_i-\mu)(x_i-\mu)^\top$
  \State $M\leftarrow \Sigma+\tfrac{1}{\lambda_{\mathrm{soft}}}I$ (take $1/\lambda_{\mathrm{soft}}=0$ if equality)
  \State \textbf{Euler/RK2:}\quad $\theta\leftarrow \theta + h\, M^{-1}\Delta$ (or a second-order variant)
\EndFor
\State \textbf{Return} the path $\{\theta(\lambda),w(\lambda)\}$
\end{algorithmic}
\end{algorithm}

\section{Theoretical Guarantees}

\begin{theorem}[Existence and uniqueness]
\label{thm:existence}
Under feasibility (Slater) and strictly positive $b$, problem \eqref{eq:primal} admits a unique optimizer. If $t\in\mathrm{relint}\,\mathrm{conv}\{\xx_i\}$ and no inequality is active at the boundary, the optimizer is strictly positive.
\end{theorem}

\begin{theorem}[Dual structure and strict concavity]
\label{thm:dual}
$L(\theta)$ in \eqref{eq:dual} is concave with $\nabla L(\theta)=t-\E_{w(\theta)}[\xx]$ and $\nabla^2 L(\theta)=-\Cov_{w(\theta)}(\xx)$. On the subspace where $\Cov_{w(\theta)}(\xx)\succ0$, $L$ is strictly concave, hence $\theta^\star$ is unique and \eqref{eq:exp-solution} yields the unique primal optimizer.
\end{theorem}

\begin{theorem}[Sensitivity]
\label{thm:sensitivity}
At the optimum, $\dfrac{\partial \theta^\star}{\partial t}=\Cov_{w^\star}(\xx)^{-1}$ and $\dfrac{\partial w^\star}{\partial t}=\mathrm{diag}(w^\star)\,(X-\1\mu^\top)\,\Cov_{w^\star}(\xx)^{-1}$ with $\mu=\E_{w^\star}[\xx]$.
\end{theorem}

\begin{theorem}[Elastic dual: strong concavity and sensitivity]
\label{thm:elastic}
$L_{\mathrm{el}}(\theta)$ is strongly concave with parameter $1/\lambda_{\mathrm{soft}}$; the maximizer is unique and Theorem~\ref{thm:elastic-sens} holds.
\end{theorem}

\begin{proposition}[Robust dual: concavity and optimality]
\label{prop:robust}
$L_{\mathrm{rob}}(\theta)=\sigma_{t_0+\mathcal U}(\theta)-\log\sum_i b_i e^{\theta^\top x_i}$ is concave. Any maximizer $\theta^\star$ yields the exponential tilt $w_i^\star\propto b_i e^{\theta^{\star\top}x_i}$. For $\mathcal U$ an $\ell_2$ ball or $\ell_\infty$ box, Algorithm~\ref{alg:prox} converges to a maximizer under standard step-size/backtracking rules (Lipschitz gradient of $f$).
\end{proposition}

\begin{theorem}[Convergence of EGMU-Newton]
\label{thm:newton}
With standard backtracking/damping, Newton ascent on $L$ is globally convergent; if $\Cov_{w(\theta)}(\xx)\succeq m I$ and $\nabla^2 L$ is Lipschitz in a neighborhood of $\theta^\star$, the rate is locally quadratic.
\end{theorem}

\begin{theorem}[Convergence of projection schemes]
\label{thm:proj}
(\emph{i}) IPF/one-dimensional KL projections cycling over equalities converge to the unique KL minimizer when feasible. 
(\emph{ii}) Bregman--Dykstra with KL distance over finitely many closed convex sets (equalities and half-spaces) converges to the KL projection onto their intersection.
\end{theorem}

\begin{remark}[Complexity]
Per Newton step: $O(NK)$ + $O(NK^2)$ to form moments and covariance, and $O(K^3)$ to solve the $K\times K$ system. 
Each 1-D projection is $O(N)$ per function/derivative evaluation (bisection/Brent). Memory footprint is $O(NK)$.
\end{remark}

\section{Implementation Notes (Stability and Scaling)}
\begin{itemize}[leftmargin=1.2em]
\item \textbf{Stability:} always use LogSumExp for partition functions; center exposures to reduce conditioning; add small ridge $\delta I$ when $\Sigma$ is nearly singular.
\item \textbf{Elastic targets (R1):} the $I/\lambda_{\mathrm{soft}}$ term improves conditioning and ensures strong concavity in the dual; recommended defaults $\lambda_{\mathrm{soft}}\in[10,10^3]$ when feasibility is uncertain.
\item \textbf{Robust sets (R2):} for $\ell_2/\ell_\infty$ sets, use Algorithm~\ref{alg:prox}; for general $\mathcal U$, combine projection oracles (or Bregman--Dykstra in $t$-space) with Moreau identity.
\item \textbf{Cap/box constraints in $w$:} half-space KL projections have 1-D solves with monotone residuals ($\frac{d}{d\lambda} a^\top w(\lambda)=-\Var_{w(\lambda)}(a)\le 0$), hence root-finding is unimodal/robust.
\item \textbf{Default solver parameters:} $\varepsilon=10^{-8}$, $c=10^{-4}$, $\beta=0.5$, $\delta=\max(10^{-10},10^{-6}\,\mathrm{tr}(\Sigma)/K)$.
\end{itemize}

\section{Extension: Multi-Period and Turnover Regularization (Brief)}
At time $t$, given previous weights $p_{t-1}$, consider
\[
\min_{w_t\in\Delta^N}\ \KL{w_t}{b}\;+\;\gamma\,\KL{w_t}{p_{t-1}}
\quad\text{s.t.}\quad X^\top w_t=\tau_t,\;A w_t\le c.
\]
This is equivalent (up to an additive constant) to $(1+\gamma)\,\KL{w_t}{\tilde b_t}$ with the \emph{effective prior}
\[
\tilde b_{t,i}\ \propto\ b_i^{\frac{1}{1+\gamma}}\;p_{t-1,i}^{\frac{\gamma}{1+\gamma}},
\]
hence the solution remains an exponential tilt $w_{t,i}\propto \tilde b_{t,i}\exp(\theta_t^\top x_i)$ and all dual/algorithmic machinery is unchanged after $b\leftarrow\tilde b_t$.
If an explicit turnover budget is desired, one may add linearized constraints or standard split variables to encode $\ell_1$-type variation limits, which fit directly into the KL-projection (Bregman--Dykstra) framework.

\section{Related Work}
\label{sec:related}
\paragraph{Information projection and exponential families.}
Our formulation is a classical $I$-projection (minimization of KL under linear moment constraints),
which yields exponential-family solutions and a concave dual with covariance Hessian; see
\citet{csiszar1975i} for the geometry of $I$-divergence, \citet{cover2006elements} for an information-theoretic treatment,
and \citet{wainwright2008graphical} for the exponential-family viewpoint connecting gradients/Hessians with moments/covariances.

\paragraph{Iterative proportional fitting and Bregman projections.}
For equality constraints, iterative proportional fitting / generalized iterative scaling (IPF/GIS)
provides a coordinate-wise Bregman projection method with convergence guarantees \citep{darroch1972generalized,csiszar1975i}.
For intersections of convex sets (equalities and half-spaces), Bregman--Dykstra cycles converge to the unique Bregman projection
onto the intersection \citep{bauschke2000dykstra}.

\paragraph{Entropy pooling and portfolio engineering.}
In portfolio applications, our setup parallels Entropy Pooling (EP), which applies cross-entropy
updating to scenario probabilities under linear ``views'' \citep{meucci2008fully}. EGMU adapts the same KL geometry to
\emph{asset weights on the simplex} with \emph{factor exposure} constraints, and makes the dual structure and sensitivity
explicitly operational for target-exposure construction.

\paragraph{Convex duality, support functions, and robustness.}
The elastic and robust variants we study are standard Fenchel--Rockafellar constructs: adding a squared penalty in the primal corresponds to
a Tikhonov (strongly concave) term in the dual; relaxing equalities to a convex target set yields a dual support function. These follow from
textbook convex analysis and duality \citep[Ch.~3--5]{boyd2004convex}, and integrate seamlessly with the exponential-family moment
structure reviewed by \citet{wainwright2008graphical}.

\paragraph{Optimization and numerical stability.}
Our damped Newton method with backtracking and ridge regularization follows standard convex-optimization practice \citep{boyd2004convex}.
Implementation details (LogSumExp stabilization, covariance centering/ridge, and moment reuse) are tailored to large-$N$, small-$K$ regimes
typical in factor construction.

\section{Conclusion}
EGMU frames target-exposure construction as KL minimization on the simplex with rigorous feasibility, uniqueness, dual structure, and sensitivity. We provide provably convergent solvers---dual Newton and KL projection (IPF/Bregman--Dykstra)---and \emph{extend} the framework to elastic/robust targets with a shared dual-moment core and furnish a path-following ODE. This yields a principled, reproducible baseline requiring minimal empirical work.

\appendix
\section*{A\quad Proofs and Technical Details}

\subsection*{A.1\quad Proof of Proposition~\ref{prop:feasible}}
Let $\mathcal{X}=\{\xx_i\}_{i=1}^N$. Since $w\in\Delta^N$ implies $X^\top w=\sum_i w_i \xx_i$, feasibility of $X^\top w=t$ is equivalent to $t\in\mathrm{conv}(\mathcal{X})$. If $t\in\mathrm{relint}\,\mathrm{conv}(\mathcal{X})$ and $b_i>0$, the KL objective is essentially smooth and strictly convex on the relative interior of the simplex, so the unique minimizer satisfies $w_i^\star>0$ by standard Lagrange multiplier/KKT arguments. With inequalities $A w\le c$, feasibility is a convex polytope; infeasibility admits a Farkas certificate (see, e.g., \citealp{boyd2004convex}, Ch.~5). \hfill$\square$

\subsection*{A.2\quad Exponential Family and Dual Structure}
Consider the Lagrangian (equalities only)
\[
\mathcal{L}(w,\lambda,\nu)=\sum_i w_i\log\frac{w_i}{b_i}+\lambda^\top (X^\top w-t)+\nu(\1^\top w-1).
\]
Stationarity in $w_i$ gives $\log w_i-\log b_i+\lambda^\top x_i+\nu+1=0$, hence
\[
w_i(\theta)=\frac{b_i e^{\theta^\top x_i}}{\sum_j b_j e^{\theta^\top x_j}},\quad \theta:=-\lambda.
\]
Substituting into the Lagrangian yields the dual $L(\theta)=\theta^\top t-\log\sum_i b_i e^{\theta^\top x_i}$. Differentiating under the softmax,
\[
\nabla L(\theta)=t-\sum_i w_i(\theta) x_i,\qquad
\nabla^2 L(\theta)=-\sum_i w_i(\theta) (x_i-\mu)(x_i-\mu)^\top=-\Cov_{w(\theta)}(x).
\]
Strict concavity holds where $\Cov_{w(\theta)}(x)\succ0$ (see \citealp{wainwright2008graphical}). \hfill$\square$

\subsection*{A.3\quad Proof of Theorem~\ref{thm:existence}}
$\KL{\cdot}{b}$ is strictly convex and lower semi-continuous on the simplex; the feasible set is convex and, under Slater, nonempty with nonempty relative interior. Hence a unique minimizer exists. Strict positivity follows from the fact that $b_i>0$ and $t\in\mathrm{relint}$ enforce finite Lagrange multipliers and thus $w_i^\star\propto b_i e^{\theta^{\star\top} x_i}>0$. \hfill$\square$

\subsection*{A.4\quad Proof of Theorem~\ref{thm:sensitivity}}
At optimum, $\nabla L(\theta^\star)=0 \iff \E_{w(\theta^\star)}[x]=t$. Differentiate both sides w.r.t.\ $t$:
\(
\frac{\partial}{\partial t}\E_{w(\theta^\star)}[x]=I.
\)
Using the exponential-family identity
\(
\frac{\partial}{\partial \theta}\E_{w(\theta)}[x]=\Cov_{w(\theta)}(x),
\)
apply the chain rule to get
\(
\Cov_{w^\star}(x)\cdot \frac{\partial \theta^\star}{\partial t}=I
\Rightarrow \frac{\partial \theta^\star}{\partial t}=\Cov_{w^\star}(x)^{-1}.
\)
For $w_i^\star=b_i \exp(\theta^{\star\top} x_i-\log Z)$, 
\[
\frac{\partial w_i^\star}{\partial \theta}
= w_i^\star (x_i-\mu)^\top,\quad \mu=\E_{w^\star}[x].
\]
Thus $\dfrac{\partial w_i^\star}{\partial t}
= w_i^\star (x_i-\mu)^\top \Cov_{w^\star}(x)^{-1}$, yielding the matrix form in the main text. \hfill$\square$

\subsection*{A.5\quad Elastic Dual and Sensitivity (Proof of Thm.~\ref{thm:elastic})}
The dual reads $L_{\mathrm{el}}(\theta)=L(\theta)-\tfrac{1}{2\lambda_{\mathrm{soft}}}\|\theta\|^2$. Hence $\nabla L_{\mathrm{el}}=\nabla L-\tfrac{1}{\lambda_{\mathrm{soft}}}\theta$ and $\nabla^2 L_{\mathrm{el}}=\nabla^2 L-\tfrac{1}{\lambda_{\mathrm{soft}}}I$, proving strong concavity. At the maximizer, $t-\E_{w(\theta)}[x]-\tfrac{1}{\lambda_{\mathrm{soft}}}\theta=0$. Differentiating w.r.t.\ $t$ and using $\partial\E_{w(\theta)}[x]/\partial\theta=\Sigma$ gives $(\Sigma+\tfrac{1}{\lambda_{\mathrm{soft}}}I)\,\partial\theta/\partial t=I$, establishing the stated sensitivities. \hfill$\square$

\subsection*{A.6\quad KL Projection onto a Single Equality (IPF step)}
Fix $u\in\Delta^N$ and the set $\mathcal{H}=\{w:a^\top w=\tau\}$. Minimize $D_{\mathrm{KL}}(w\|u)$ subject to $a^\top w=\tau$ and $\1^\top w=1$. Stationarity: $\log(w_i/u_i)+1+\alpha a_i+\nu=0$, so $w_i\propto u_i e^{\alpha a_i}$. The normalization ensures $w(\alpha)\in\Delta^N$. Define $\phi(\alpha)=a^\top w(\alpha)-\tau$. One computes $\phi'(\alpha)=\Var_{w(\alpha)}(a)>0$ unless $a$ is degenerate, hence a unique root exists and can be found by bisection. \hfill$\square$

\subsection*{A.7\quad KL Projection onto a Half-space (Inequality step)}
For $\mathcal{H}=\{w: a^\top w\le \tau\}$, if $u$ is feasible, the projection is $u$. Otherwise, the KKT conditions yield $w(\lambda)\propto u\odot e^{-\lambda a}$ with $\lambda\ge0$ chosen so that $a^\top w(\lambda)=\tau$. Monotonicity follows from $\frac{d}{d\lambda} a^\top w(\lambda)=-\Var_{w(\lambda)}(a)\le0$. \hfill$\square$

\subsection*{A.8\quad Convergence of EGMU-Newton (Refinement of Thm.~\ref{thm:newton})}
The objective $L(\theta)=\theta^\top t-\log\sum_i b_i e^{\theta^\top x_i}$ is twice continuously differentiable and concave, with $\nabla L(\theta)=t-\E_{w(\theta)}[x]$ and $\nabla^2 L(\theta)=-\Cov_{w(\theta)}(x)$. If $\|x_i\|_2\le R$ for all $i$, then $\|\nabla^2 L(\theta)\|\le R^2$ for all $\theta$, and $\nabla^2 L$ is locally Lipschitz (with constant depending on $R$ and the third centered moment). Under these mild smoothness conditions, damped Newton with Armijo backtracking is globally convergent and locally quadratically convergent in a neighborhood of $\theta^\star$ for strongly concave $L$ on the relevant subspace (see \citealp{boyd2004convex}, Ch.~9). Ridge regularization $(\Sigma+\delta I)$ stabilizes solves when $\Sigma$ is ill-conditioned; as $\delta\downarrow 0$ the step approaches the exact Newton direction.

\subsection*{A.9\quad Convergence of IPF and Bregman--Dykstra (Proof of Thm.~\ref{thm:proj})}
Part (i) follows from Csisz\'ar's $I$-projection theory and the Darroch--Ratcliff analysis of generalized iterative scaling for log-linear models \citep{csiszar1975i,darroch1972generalized}. Part (ii) is a special case of Dykstra's algorithm with Bregman divergences: for finitely many closed convex sets and a Legendre-type Bregman generator (negative entropy here), the cyclic projections converge to the unique Bregman projection onto the intersection \citep{bauschke2000dykstra}. \hfill$\square$

\subsection*{A.10\quad Carath\'eodory support bound (remark)}
Any $t\in\mathrm{conv}\{\xx_i\}$ admits a representation using at most $K+1$ points. See, e.g., \citet{barvinok2002convex}. This yields a sparsity upper bound for exact feasibility, though KL minimization under strictly positive prior typically produces dense solutions unless boundary constraints are active.

\subsection*{A.11\quad Robust dual and proximal map (details)}
Let $g(y)=\iota_{\,t_0+\mathcal U}(y)$. Its Fenchel conjugate is
$g^*(\theta)=\sup_{y}\{\theta^\top y-g(y)\}=\sup_{u\in\mathcal U}\theta^\top(t_0+u)=\theta^\top t_0+\sigma_{\mathcal U}(\theta)$,
hence the robust dual in \S\ref{sec:robust}. For the proximal step, use Moreau's identity for conjugates:
$\mathrm{prox}_{\eta g^*}(z)=z-\eta\,\mathrm{prox}_{g/\eta}(z/\eta)$.
Since $g/\eta$ is the indicator of $t_0+\mathcal U$, $\mathrm{prox}_{g/\eta}(z/\eta)=\Pi_{\,t_0+\mathcal U}(z/\eta)$.
With the translation $y\mapsto y-t_0$, this yields
$\mathrm{prox}_{\eta\,\sigma_{\mathcal U}}(z)=z-\eta\,\Pi_{\mathcal U}(z/\eta)$ used in Algorithm~\ref{alg:prox}. \hfill$\square$

\subsection*{A.12\quad Existence and uniqueness of the solution path ODE}
For $t(\lambda)=t_0+\lambda\Delta$, the optimal $\theta(\lambda)$ satisfies
$F(\theta,\lambda):=t(\lambda)-\E_{w(\theta)}[x]-\tfrac{1}{\lambda_{\mathrm{soft}}}\theta=0$.
Then $\partial_\theta F(\theta,\lambda)=\Sigma(\theta)+\tfrac{1}{\lambda_{\mathrm{soft}}}I\succeq mI$ on a neighborhood where
$\Sigma$ is bounded below. By the implicit function theorem, there exists a unique $C^1$ path $\theta(\lambda)$ with
$\dfrac{d\theta}{d\lambda}=\big(\Sigma(\theta)+\tfrac{1}{\lambda_{\mathrm{soft}}}I\big)^{-1}\Delta$.
Under locally Lipschitz $\nabla^2 L$, Euler and RK2 integrators achieve $O(h)$ and $O(h^2)$ global errors respectively. \hfill$\square$

\section*{Classification and availability}
\textbf{JEL:} G11, C61, C63, C58.\quad
\textbf{MSC 2020:} 90C25, 90C90, 62F10, 94A17.\quad
\textbf{Reproducibility:} Minimal synthetic scripts (Newton/IPF/ProxGrad/Path) to reproduce algorithms and figures are provided in the supplementary material; no proprietary data are used.

\bibliographystyle{abbrvnat}
\bibliography{references}
\end{document}